\documentclass[epj]{svjour}
\usepackage{graphics}
\usepackage{rotating}
\begin{document}
\title{Further considerations concerning claims for deeply bound kaon atoms
 and reply to criticisms}
\author{E. Oset\inst{1} 
\and V.K Magas\inst{2}
\and A. Ramos\inst{2}
\and H. Toki\inst{3}
}                     
%
%
\institute{Departamento de F\'{\i}sica Te\'orica and IFIC,
Centro Mixto Universidad de Valencia-CSIC, 
Institutos de
Investigaci\'on de Paterna, Aptd. 22085, 46071 Valencia, Spain \and 
Departament d'Estructura i Constituents de la Materia, Universitat de Barcelona,
Diagonal 647, 08028 Barcelona, Spain\and
Research Center for Nuclear Physics, Osaka University,
Ibaraki, Osaka 567-0047, Japan  
} 
\date{Received: date / Revised version: date}
%

\abstract{ We briefly review the experiments of KEK and FINUDA, that
claim evidence for deeply bound kaon states, from the perspective of recent theoretical
papers and experiments that provide an alternative explanation of the peaks seen. 
At the same time we show that recent criticisms raised by
Akaishi and Yamazaki, and exposed by Akaishi in this Conference, have no base. 
}
%

\maketitle
\section{Introduction}
\label{intro}
A brief story of the recent events around deeply bound kaons atoms could be made
as follows: Chiral potentials \cite{lutz,angelsself,schaffner,cieply,tolos}
provide potentials of depth around 50 MeV attraction at a width around 100 MeV at 
$\rho =\rho_0$. With these potentials, deeply bound states of binding energy around
30-40 MeV are obtained, but with a width of the order of 100 MeV which would 
preclude the observation of peaks \cite{okumura}. A next step of the development
appeared with the claims of a very large attractive potential in light nuclei in
\cite{akaishi,akainew} (AY), around 650 MeV at the center of the nucleus in 
\cite{akainew} and with the matter compressed to 10 times nuclear matter
density.  An experiment was made at KEK with the $K^-$ absorption at rest in 
$^4He$  \cite{suzuki} and a peak was seen and attributed first to a strange 
tribaryon, since its interpretation as a deeply kaon bound state would 
contradict the predictions of \cite{akaishi}, but afterwords it was reinterpreted 
as a deeply bound kaon atom since it would match with the corrected version of
the potential in \cite{akainew}. The FINUDA collaboration made the same
experiment in different nuclei and found a broad peak in the $\Lambda p$ back
to back invariant mass which was attributed to the existence of the $K^- pp$ bound
state  \cite{finuda}. With the interpretation of these peaks in clear
contradiction with the predictions of the chiral potentials, theoreticians come
into the scene: Oset and Toki (OT) \cite{prc} write a paper indicating that the peak seen 
at KEK is no proof of a kaon bound state since it can be interpreted in terms 
of absorption of the $K^-$ by a pair of nucleons going to $\Sigma p$,
  with the daughter nucleus 
left as spectator. Parallely, Oset, Magas, Ramos and Toki (MORT) write a paper
\cite{magas} and provide an alternative explanation of the peak seen at FINUDA 
as coming from
$K^- $ absorption in the nucleus going to $\Lambda N$, followed by the 
rescattering of the $\Lambda$ or the  nucleon with the daughter nucleus. 
  After that, a KEK like experiment is made at FINUDA  looking at proton spectra
following $K^-$ absorption at rest and a peak is indeed found in $^6 Li$ 
\cite{newfinuda} which,
thanks to the measurement of pions in coincidence, allows the authors to 
interpret it as coming from $K^-$ absorption by a pair of nucleons 
going to $\Sigma p$, with the daughter nucleus left as spectator, just the
explanation offered by OT in \cite{prc} for the KEK peak. Incidentally, a second peak
seen in the KEK experiment when making a cut of slow pions, and attributed to 
 $K^-$ absorption by a pair of nucleons going to $\Lambda p$ in \cite{prc} is
 also seen in \cite{newfinuda} as a feeble signal and associated to the 
 $\Lambda p$ mechanisms as suggested in \cite{prc}.  This peak is, however, much
 better seen, as a very narrow peak, in the the $\Lambda p$ back to back 
 invariant mass spectrum of the FINUDA experiment \cite{finuda}.
 
 In between, two novelties
 have appeared from the japanese side, the experiment has been redone, with an
 inclusive measurement, without the cuts and acceptance of \cite{suzuki}, and the
 peaks seem to disappear as reported by Iwasaki in this Conference
 \cite{iwasaki}. It was,
 however, indicated in the discussion that the useful measurement is the one with
 cuts that reduces background and stresses peaks, which has not been redone 
 \cite{iwaprivate}, and that the the FINUDA data
 on the proton spectrum showing the KEK like peaks is there to be also seriously
 considered. The other novelty is the paper by Akaishi and Yamazaki 
 \cite{criti} criticizing
 both the approaches of \cite{prc} and \cite{magas}, and the extra criticism of
 Akaishi in this Conference criticizing the chiral approach, because of the 
 "unrealistic  range" of the interaction used. Actually, no range is used for the
 interaction because, as we shall see below, all recent versions of the
 chiral approach rely upon the N/D method and dispersion relations, which only
 requires the knowledge of the interaction on shell, and the range of the
 interaction never appears in the formalism. In what follows we show that the
 recent criticism of Akaishi and Yamazaki, in \cite{criti} and of Akaishi in his
 talk have no base.

\section{The chiral approach and the N/D method}
\label{sec:1}
The chiral approach of \cite{angels}, with the on-shell factorization of the 
potential and the t-matrix, is based on the N/D method.
One can find a systematic and easily comprehensible derivation 
 of the  ideas of the N/D method applied for the first time to the meson baryon system in
 \cite{ulfnsd}, which we reproduce here below and which follows closely
 the similar developments used before in the meson-meson interaction \cite{ollernsd}.
 One defines the transition $T-$matrix as $T_{i,j}$ between the coupled channels which couple to
 certain quantum numbers. For instance in the case of  $\bar{K} N$ scattering studied in
 \cite{ulfnsd} the channels with zero charge are $K^- p$, $\bar{K^0} n$, $\pi^0 \Sigma^0$,$\pi^+
 \Sigma^-$, $\pi^- \Sigma^+$, $\pi^0 \Lambda$, $\eta \Lambda$, $\eta \Sigma^0$, 
 $K^+ \Xi^-$, $K^0 \Xi^0$.
 Unitarity in coupled channels is written as
 
\begin{equation} 
Im T_{i,j} = T_{i,l} \rho_l T^*_{l,j}
\end{equation}
where $\rho_i \equiv 2M_l q_i/(8\pi W)$, with $q_i$  the modulus of the c.m. 
three--momentum, and the subscripts $i$ and $j$ refer to the physical channels. 
 This equation is most efficiently written in terms of the inverse amplitude as
\begin{equation}
\label{uni}
\hbox{Im}~T^{-1}(W)_{ij}=-\rho(W)_i \delta_{ij}~,
\end{equation}
The unitarity relation in Eq. (\ref{uni}) gives rise to a cut in the
$T$--matrix of partial wave amplitudes, which is usually called the unitarity or right--hand 
cut. Hence one can write down a dispersion relation for $T^{-1}(W)$ 
\begin{equation}
\label{dis}
T^{-1}(W)_{ij}=-\delta_{ij}~g(s)_i+ V^{-1}(W)_{ij} ~,
\end{equation}
with
\begin{equation}
\label{g}
g(s)_i=\widetilde{a}_i(s_0)+ \frac{s-s_0}{\pi}\int_{s_{i}}^\infty ds' 
\frac{\rho(s')_i}{(s'-s)(s'-s_0)},
\end{equation}

where $s_i$ is the value of the $s$ variable at the threshold of channel $i$ and 
$V^{-1}(W)_{ij}$ indicates other contributions coming from local and 
pole terms, as well as crossed channel dynamics but {\it without} 
right--hand cut. These extra terms
are taken directly from $\chi PT$ 
after requiring the {\em matching} of the general result to the $\chi PT$ expressions. 
Notice also that $g(s)_i$ is the familiar scalar loop integral of a meson and a
baryon propagators.

One can simplify the notation by employing a matrix formalism. 
Introducing the 
matrices $g(s)={\rm diag}~(g(s)_i)$, $T$ and $V$, the latter defined in 
terms 
of the matrix elements $T_{ij}$ and $V_{ij}$, the $T$-matrix can be written as:
\begin{equation}
\label{t}
T(W)=\left[I-V(W)\cdot g(s) \right]^{-1}\cdot V(W)~.
\end{equation}
which can be recast in a more familiar form as 
 \begin{equation}
\label{ta}
T(W)=V(W)+V(W) g(s) T(W)
\end{equation}
Now imagine one is taking the lowest order chiral amplitude for the kernel
$V$ as done in
\cite{ulfnsd}. Then the former equation is nothing but the Bethe Salpeter equation with the
kernel taken from the lowest order Lagrangian and  factorized  on  shell, the same
approach followed in \cite{angels}, where different arguments were used to justify the on shell
factorization of the kernel. The kernel V plays the role of a potential in
ordinary Quantum Mechanics.

The on shell factorization of the kernel, justified here with the N/D method,
renders the set of coupled Bethe Salpeter integral equations a simple set of
algebraic equations.

  The important thing to note is that both the kernel and the $T$ matrix only
  appear on shell, for a value of $\sqrt{s}$. The range of the interaction is
  never used. The loop function is made convergent via a subtraction in the
  dispersion relation, or equivalently a cut off in  the three momentum as used
  in \cite{angels}, which is proved to be equivalent to the subtraction method
  in \cite{ulfnsd}.  Akaishi in his talk confuses this cut off in the loop of
  propagators with the range of the interaction, when they have nothing to do
  with each other. Even more, the theory must be cut off independent, which
  means, one can change arbitrarily the cut off by introducing the appropriate
  higher order counterterms. As a consequence of this, all pathologies of the
  interaction pointed out by Akaishi in his talk are a pure invention, which has
  nothing to do with the physics of the problem.

\section{Interpretation of the narrow  FINUDA peaks and KEK peaks}
\begin{figure}[h]
\begin{center}
\begin{turn}{-90}
  \includegraphics[width=0.3\textwidth]{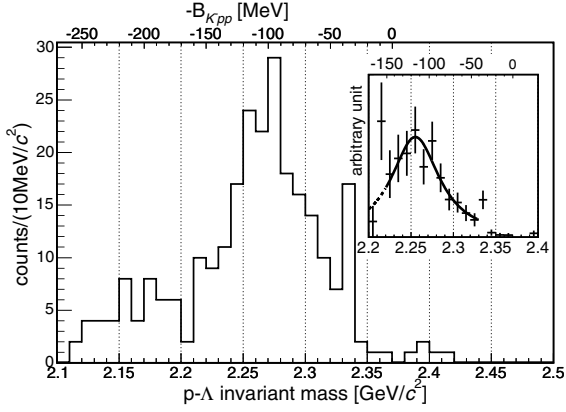}
\end{turn}
\caption{$\Lambda p$ invariant mass distribution of back to back pairs 
following $K^-$ absorption in a mixture of nuclei, $^6 Li$, $^7 Li$ and
$^{12}C$. The inset of the figure shows data corrected for the 
detector acceptance. From \cite{finuda}.}
\label{fig:1}
\end{center}
\end{figure}

In \cite{finuda}, for absorption in a sample of $^6 Li$, $^7 Li$, $^{12}C$, a
narrow peak is seen at $M_I =2340 MeV$ of the back to back $\Lambda p$ system 
and a wider one at $M_I =2275 MeV$, see  Fig. ~\ref{fig:1}.
Let us assume $^7Li$ for simplicity of the discussion.
 The first thing to recall
is the experience of pion absorption that concluded that at low pion energies
the absorption was dominated by a direct two body process (even if later on
there would be rescattering of the nucleons in the nucleus, giving rise to what
was called indirect three body absorption in contrast with the possible direct
three body absorption which had a small rate at low energies \cite{weyer}.)
We consider the $K^-$ two nucleon absorption mechanism, disregarding the one body
mechanisms which do not produce $\Lambda p$ back to back, see
Fig.~\ref{fig:1}.\\

{\bf Origin of the narrow peak at $M_I =2340 MeV$} :

We have the reaction,

\begin{equation}
 K^-  pp  + (^5H spectator) \to  \Lambda p + (^5H spectator)
\end{equation}

The kinematics of the reaction is as follows: Let $P$ be the total momentum of
the $K¯$-nucleus system, and 
 $p_1$, $p_2$ and $p_3$ the momenta of the $\Lambda$, $p$ and 
$^5H$ spectator respectively. We have

\begin{equation}
(P-p_3)^2= (p_1 +p_2 )^2= {M_{12}}^2 
\end{equation}
from where we deduce that 
\begin{equation}
\Delta(M_{12})=M(K~Li)\Delta(E_3)/M_{12}
\end{equation}

This would lead to $\Delta(M_{12})\sim 10 MeV$ for absorption in $^4He$ and 
$\Delta(M_{12})\sim 1 MeV$ for $^7Li$ if one takes as representative of the Fermi
momentum of the quasideuteron or $pp$ pairs $150~ MeV$ for  $^4He$ and $50 ~MeV $
for $^7Li$ as suggested in \cite{criti}.  This produces a dispersion of the
$p$ momentum in the CM of the same order of magnitude. This quantity is smaller
than the main source for $p$ momentum dispersion which is the boost of the
proton from the CM of $\Lambda p$ to the frame where the $\Lambda p$ has the Fermi
momentum of the initial NN pair, $p_{NN}$.

The boost is easily implemented requiring only nonrelativistic kinematics. We
have

\begin{equation}
\vec{p_p}=\vec{p_{CM}} + m_p~ \vec{V} ;~~~~ \vec{V}=\vec{p_{NN}}/M_{12}
\end{equation}
\begin{equation}
\Delta(\vec{p_p})^2=m_p^2 ~\vec{V}^2 /3
\end{equation}
\begin{equation}
\Delta(p_p)= \pm 35 ~MeV/c ~(11~MeV/c)
\end{equation}
\begin{equation}
 ~~~~~~~~~~~~~~~~~~for~ p_{NN}=150 ~ MeV/c ~(50 ~MeV/c)
\end{equation}

\begin{figure}[h]
\begin{center}
\begin{turn}{-90}
  \includegraphics[width=0.35\textwidth]{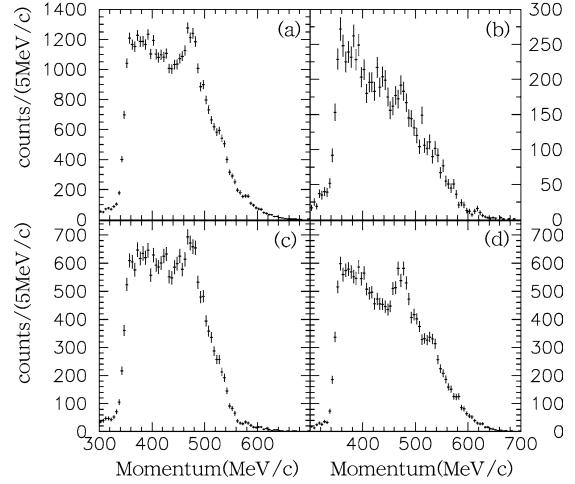}
\end{turn}
\caption{Proton spectra following $K^-$ absorption in $^4 He$. Lower two
figures: left with high pion momentum cut, right with lower pion momentum cut.
From \cite{suzuki}.}
\label{fig:2}
\end{center}
\end{figure}

Hence we would have a dispersion of proton momentum of $\pm 35 ~MeV$ for $K^-$
absorption in $^4 He$ and $\pm 11 ~MeV$ in the $^7Li$ case. The exercise has
been done for $K^- pp \to \Lambda p$ but the results are the same if one has 
$K^- NN \to \Sigma p$. This latter reaction was the one suggested by OT in
\cite{prc} to explain the KEK peak seen in Fig.~\ref{fig:2}, lower left figure
of the panel, around $475 ~ MeV$.  The dispersion of Eq. (12) 
would roughly agree with the peak.

We should note that the peaks can be made more narrow, as we have checked
numerically by: 1) assuming absorption from a $2p$ orbit of the $K^-$, 
2) forcing the $\Lambda p$ pair to go back
to back, 3) putting restrictions on the pion momenta.   

\begin{figure}[h]
\begin{center}
\begin{turn}{-90}
  \includegraphics[width=0.3\textwidth]{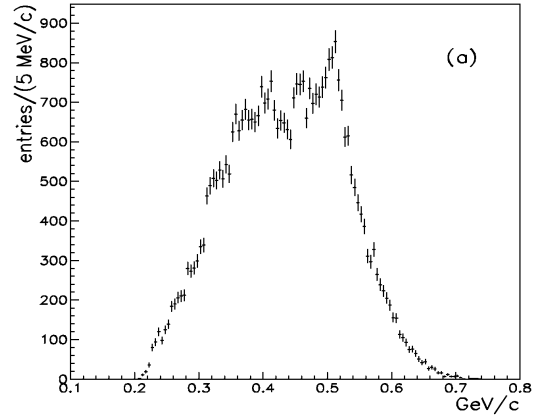}
\end{turn}
\caption{Proton momentum distribution following the absorption of $K^-$ in $^6
Li$ from \cite{newfinuda}.}
\label{fig:3}
\end{center}
\end{figure}

It is interesting to
observe in this respect that in the figure of the KEK experiment in the case
when the slow pions are selected (lower right figure in the panel ) one can see
also a peak in the momentum distribution  at $p\sim 545 MeV$, which was 
identified in
\cite{prc} as coming from $K^-$ absorption going to   $\Lambda p$ 
 with the daughter nucleus as a spectator. It is interesting to see that such a
 signal, "a feeble signal around 580 MeV/c" is seen even in the inclusive 
 spectrum of \cite{newfinuda}, see Fig.~\ref{fig:3} , and correctly identified there as coming
 from $K^-$ absorption in $^6Li$ going to $\Lambda p$ (note one has smaller binding of
 the nucleons here than one has in $^4 He$ and there is no loss of energy as in
 the case of a thick target of  \cite{suzuki}).
 
Coming back to the absorption of $K^-$ in $^4He$ it should be noted that the
candidate reaction for the peak at 475 MeV/c is the reaction with the rate 
\begin{equation}
 \Sigma^- p~d ~~~~~1.6 ~\%
\end{equation}

which has been measured by \cite{katz}. A fraction of this reaction can go with
the $d$ as a spectator, and then it is worth mentioning that the fraction of the
cross section of this peak is estimated in \cite{suzuki} at less than 1\%,  of
the order of 0.34 \% according to \cite{thesis}.

\section{Interpretation of the wide FINUDA peak}

\begin{figure}[h]
\begin{center}
  \includegraphics[width=0.5\textwidth]{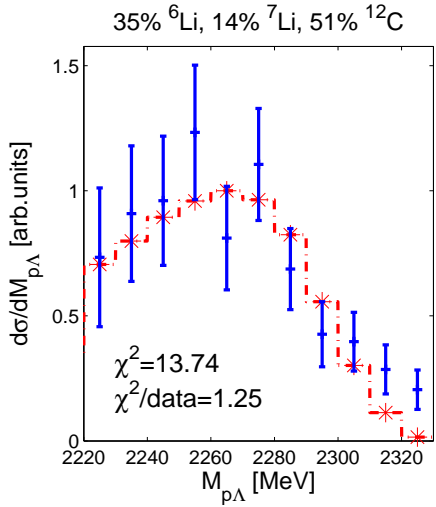}
\end{center}
\caption{Theoretical calculation of \cite{magas} versus experiment of the $\Lambda p$ 
invariant mass distribution of back to back pairs 
following $K^-$ absorption in a mixture of nuclei, $^6 Li$, $^7 Li$ and
$^{12}C$. Histogram theory, bars data from \cite{finuda}.}
\label{fig:4}
\end{figure}

Next we turn to the wide FINUDA peak in the experiment \cite{finuda}. This peak
was interpreted as naturally coming from the absorption of a $K^-$ from the
nucleus going to $\Lambda N$ followed by a rescattering of the nucleon or the
Lambda with the remnant nucleus \cite{magas}. This is the equivalent of the quasielastic peak
which appears in all inclusive nuclear reactions with a similar width which is
due to the Fermi motion of the  nucleons.  In \cite{magas} a calculation was
done for the mixture of the different nuclei, as in the experiment and the
results are seen in  Fig.~ \ref{fig:4}

\section{Claims of no peaks in Akaishi and Yamazaki}

Another of the points in the work of \cite{criti} is that the peaks predicted by
OT in \cite{prc} and MORT in \cite{magas} are unrealistic and that a proper
calculation does not produce any peaks. The curious results are a consequence of
a calculation in \cite{criti} that:

1) Considers absorption by all four particles at once in the $^4He$ case,
disregarding the dynamics found from pion absorption.
The spectra essentially reflect phase space with four particles in the final 
state .

2) Does not consider absorption by two N with spectator remnant nucleus.

3) Does not consider angular cuts or particles in coincidence.

4) Does not consider rescattering of particles.\\

And with all this an obvious broad  spectrum is obtained.

Consequently with their finding the authors of \cite{criti} write textually: 

"OT further insist that the same $K^-NN$ absorption mechanism at rest 
persists in the case of heavier targets as well ($^7$Li and $^{12}$C). 
However, this proposal contradicts the FINUDA experiment \cite{finuda}, 
in which they reconstructed an invariant mass spectrum of 
$M_{\rm inv}(\Lambda p)$. 
  Contrary to the naive expectation of OT, the spectrum shows no such 
  peak at 2340 MeV/$c^2$. "
  
  This assertion could not be more illuminating of the criticism raised. The
  peak that Akaishi and Yamazaki claim that we predict and does not exist is
  the one seen exactly at 2340 MeV/$c^2$ in Fig. 3 of \cite{finuda},
  which we have reproduced in  Fig. ~\ref{fig:1}.
  
  In case there could be some doubts about this peak let us quote textually what
  the authors of \cite{finuda} say regarding this peak:
  
  "On the other hand, the detector system is very sensitive to the existence of
  the two nucleon  absorption mode $K^- + "pp" \to \Lambda p$ since its
  invariant mass resolution is 10 MeV/$c^2$ FWHM. The effect of the nuclear
  binding of two protons is only to move the peak position to the lower mass
  side of the order of separation energies of two protons ($\sim$ ~30 MeV), and
  not to broaden the peak. A sharp spike around 2.34 GeV/$c^2$ may be
  attributed to this process."
  
  Incidentally this mechanism is the one proposed by \cite{prc} to explain the
  peak at 545 MeV/c in the proton spectrum when the slow pion cut is made in
  \cite{suzuki}. This peak is also the one that shows in the inclusive 
  momentum spectrum of \cite{newfinuda} mentioned there as a "feeble signal " and
  with the same interpretation.
  
  With their claims than no peaks should be seen from these processes AY
  obviously also contradict the clear peak seen in  \cite{newfinuda} around 
  500 MeV/c, see Fig.~\ref{fig:3}, which, with the detection in coincidence of the
  pions coming from $\Sigma \to \pi N$ decay, the authors of \cite{newfinuda}
  unmistakeably relate to the $K^- NN \to \Sigma N$ process, the mechanism
  proposed by OT to explain the lower peak of the spectrum in \cite{suzuki}.

  Another of the "proofs" presented in \cite{criti} is a spectrum of 
  $K^- ~^4He \to \Lambda ~ d~ n$ in which no peak around 560 MeV/c is seen, as
  one could guess from our interpretation of the process.  The comparison is,
  however, inappropriate. First, the number of counts is of the order of three counts per
  bin, as average. Second, the rate of  $K^-~^4He  \to \Sigma^- p~d$ 
  of  1.6 ~\% according
  to \cite{katz}, and the rate of the peak of the KEK experiment that we
  attribute to this process, with a value of the order of 0.3 \%, indicate that
  only a fraction of this process will go with the $d$ as a spectator, leading
  to a peak that can only be seen with  far better statistics and
  resolution than the one in the spectrum of the $K^- ~^4He \to \Lambda ~ d~ n$ 
  experiment \cite{katz}.

  Finally, AY present another calculation to prove that the broad FINUDA
  invariant mass peak requires an explanation based on the $K^- pp$ bound system
  by 115 MeV.  Their results are presented in Fig. 7 of their paper. 
  The calculation made is, however, simply unacceptable. They make
  the following assumptions:

1) Calculation in $^4He$ and compare to experiment which is a mixture of 
$^6Li,~ ^7Li,~ ^{12}C$.

2) Direct absorption by four nucleons.

3) No dynamics, just phase space.

4) Has no rescattering, shown by Magas to be essential to account 
for the peak.

And with this calculation they claim that the $K^- pp$ cluster is
 bound by 115 MeV !!!. 

  We should also mention here another example of inappropriate comparison. In
  Fig. 7 of \cite{criti}, which aims at describing the wide FINUDA peak, a
  vertical line is plotted with the lable OT, presenting this as the position 
 predicted by OT for this FINUDA peak. This comparison is out of place because
 OT in \cite{prc} never attempted to predict this broad FINUDA peak. This is
 done by MORT in \cite{magas}, requiring a different mechanism, the rescattering
 of the proton or the $\Lambda$ after $K^-$ absorption by two nucleons 
  \cite{finuda}, which automatically produces a peak at lower invariant masses.

\section{conclusions}
\begin{itemize}
\item Akaishi and Yamazaki criticisms of Oset Toki and Magas et al, are unfounded.

\item AY potential with 10 $\rho_0$  compressed matter should not be considered serious.

\item The claims of KEK and FINUDA for deeply bound kaons were unfounded.

\item The new FINUDA data on $p$ spectrum following $K^-$ absorption in $^6Li$
has been very clarifying, showing KEK like peaks and interpreting them with
the  suggestion of Oset and Toki.

\item The new calculations of Dote and Weise \cite{weise}, and

 Schevchenko, 
Gal, Mares \cite{shevchenko} predicting a bound 

$K^- pp$ state with 50-70 MeV
binding, but more that 100 MeV width, have  brought new light to this issue.
They do not support the deeply bound narrow $K^- pp$ systems claimed by FINUDA.

\item The new measurements of $^4He$ X rays by Hayano, Iwasaki et al.
\cite{hayano} are very important 
to clarify the issue. They clearly contradict predictions of Akaishi based on
his potential. 

\item Interesting results from COSY, Buescher et al from 
$p ~d \to K^+ K^- ~^3He$ in the 
same direction \cite{grishina}, clearly rejecting such large $K^-~ ^3He$ potentials. 
 
\end{itemize}

\section{Aknowledgments} This work is 
partly supported by DGICYT contract number BFM2003-00856, the
Generalitat Valenciana
and the E.U. EURIDICE network contract no. HPRN-CT-2002-00311.
This research is part of the EU Integrated Infrastructure Initiative
Hadron Physics Project under contract number RII3-CT-2004-506078.


\begin{thebibliography}{99}


  
\bibitem{lutz}
  M.~Lutz,
  Phys.\ Lett.\ B {\bf 426} (1998) 12
  [arXiv:nucl-th/9709073].
  
\bibitem{angelsself}
  A.~Ramos and E.~Oset,
  Nucl.\ Phys.\ A {\bf 671} (2000) 481
  [arXiv:nucl-th/9906016].
  
\bibitem{schaffner}
  J.~Schaffner-Bielich, V.~Koch and M.~Effenberger,
  Nucl.\ Phys.\ A {\bf 669} (2000) 153
  [arXiv:nucl-th/9907095].
  
\bibitem{cieply}
  A.~Cieply, E.~Friedman, A.~Gal and J.~Mares,
  Nucl.\ Phys.\ A {\bf 696} (2001) 173
  [arXiv:nucl-th/0104087].
  
\bibitem{tolos}
  L.~Tolos, A.~Ramos and E.~Oset,
  Phys.\ Rev.\ C {\bf 74} (2006) 015203
  [arXiv:nucl-th/0603033].

  
\bibitem{okumura}
  S.~Hirenzaki, Y.~Okumura, H.~Toki, E.~Oset and A.~Ramos,
  Phys.\ Rev.\ C {\bf 61} (2000) 055205.
  
\bibitem{akaishi}
  Y.~Akaishi and T.~Yamazaki,
  Phys.\ Rev.\ C {\bf 65} (2002) 044005.


  
\bibitem{akainew}
  Y.~Akaishi, A.~Dote and T.~Yamazaki,
  Phys.\ Lett.\ B {\bf 613} (2005) 140
  [arXiv:nucl-th/0501040].
  
\bibitem{suzuki}
  T.~Suzuki {\it et al.},
  Phys.\ Lett.\ B {\bf 597} (2004) 263.
  
\bibitem{finuda}
  M.~Agnello {\it et al.}  [FINUDA Collaboration],
  Phys.\ Rev.\ Lett.\  {\bf 94} (2005) 212303.
  
\bibitem{prc}
  E.~Oset and H.~Toki,
  Phys.\ Rev.\ C {\bf 74} (2006) 015207
  [arXiv:nucl-th/0509048].
  
\bibitem{magas}
  V.~K.~Magas, E.~Oset, A.~Ramos and H.~Toki,
  Phys.\ Rev.\ C {\bf 74} (2006) 025206
  [arXiv:nucl-th/0601013]; arXiv:nucl-th/0611098.
  
\bibitem{newfinuda}
  M.~Agnello {\it et al.}  [FINUDA Collaboration],
  Nucl.\ Phys.\ A {\bf 775} (2006) 35.
  
\bibitem{iwasaki} M. Iwasaki, talk in this Conference.

\bibitem{iwaprivate} From public discussion and M. Iwasaki, private communication.
  
  
\bibitem{criti}
  T.~Yamazaki and Y.~Akaishi,
  arXiv:nucl-ex/0609041.
  
\bibitem{angels}
  E.~Oset and A.~Ramos,
  Nucl.\ Phys.\ A {\bf 635} (1998) 99
  [arXiv:nucl-th/9711022].

 
\bibitem{ulfnsd}
  J.~A.~Oller and U.~G.~Meissner,
  Phys.\ Lett.\ B {\bf 500} (2001) 263
  [arXiv:hep-ph/0011146].
 
\bibitem{ollernsd}
  J.~A.~Oller and E.~Oset,
  Phys.\ Rev.\ D {\bf 60} (1999) 074023
  [arXiv:hep-ph/9809337].
  

 
\bibitem{weyer}
  A.~Lehmann {\it et al.}  [LADS Collaboration],
  Phys.\ Rev.\ C {\bf 55} (1997) 2931.
  
\bibitem{twolamb}
  V.~K.~Magas, E.~Oset and A.~Ramos,
  Phys.\ Rev.\ Lett.\  {\bf 95} (2005) 052301
  [arXiv:hep-ph/0503043].
  
\bibitem{katz}
  P.~A.~Katz, K.~Bunnell, M.~Derrick, T.~Fields, L.~G.~Hyman and G.~Keyes,
  Phys.\ Rev.\ D {\bf 1} (1970) 1267.
  
 \bibitem{thesis} T. Suzuki, PhD Thesis and H. Outa, private communication.
 \bibitem{weise} W. Weise in this Conference and A. Dote in this Conference.
 
\bibitem{shevchenko}
  N.~V.~Shevchenko, A.~Gal and J.~Mares,
  arXiv:nucl-th/0610022.
  
  \bibitem{hayano} R. Hayano, talk in this Conference.

\bibitem{grishina}
  V.~Y.~Grishina, M.~Buscher and L.~A.~Kondratyuk,
  arXiv:nucl-th/0608072.

\end{thebibliography}
%

\end{document}